\documentclass[twocolumn,english]{revtex4-2}
\usepackage[T1]{fontenc}
\usepackage[latin9]{inputenc}
\usepackage{amsmath}
\usepackage{amssymb}
\usepackage{graphicx}
\usepackage{babel}
\begin{document}
\global\long\def\cm{\text{cm}^{-1}}%
\global\long\def\dd{\text{D}_{2}}%
\global\long\def\mdd{\text{mD}_{2}}%

\title{Thermalization of open quantum systems using multiple-Davydov $\text{D}_{2}$
variational approach}
\author{Mantas Jaku\v{c}ionis, Darius Abramavi\v{c}ius\textsuperscript{}}
\affiliation{\textsuperscript{}Institute of Chemical Physics, Vilnius University,
Sauletekio Ave. 9-III, LT-10222 Vilnius, Lithuania}
\begin{abstract}
Numerical implementations of explicit phonon bath require a large
number of oscillator modes in order to maintain oscillators at the
initial temperature when modelling energy relaxation processes. An
additional thermalization algorithm may be useful in controlling the
local temperature. In this paper we extend our previously proposed
thermalization algorithm \citep{Jakucionis2021} to be used with the
numerically exact multiple-Davydov $\text{D}_{2}$ trial wavefunction
for simulation of relaxation dynamics and spectroscopic signals of
open quantum systems using the time-dependent Dirac-Frenkel variational
principle. By applying it to the molecular aggregate model, we demonstrate
how the thermalization approach significantly reduces the numerical
cost of simulations by decreasing the number of oscillators needed
to explicitly simulate the aggregate's environment fluctuations, while
maintaining correspondence to the exact population relaxation dynamics.
Additionally, we show how the thermalization can be used to find the
equilibrium state of the excited molecular aggregate, which is necessary
for simulation of the fluorescence and other spectroscopic signals.
The presented thermalization algorithm open possibilities to investigate
larger system-bath models than was previously possible using the multiple-Davydov
$\text{D}_{2}$ trial wavefunction and local heating effects in molecular
complexes.
\end{abstract}
\maketitle

\section{Introduction}

Open quantum system models are widely used to describe properties
of molecular aggregates \citep{Mukamel1995,Valkunas2013a}. The \textit{system}
usually consists of molecular electronic states. Intramolecular vibrational
degrees of freedom (DOFs), which play the major role in relaxation
process of the systems of interest can also be included into the quantum
system model. The rest of the DOFs are treated as an environment of
a constant temperature -- the \textit{bath}. The bath is modelled
as a collection of quantum harmonic oscillators (QHO) and is characterized
by a continuous fluctuation spectral density function \citep{Valkunas2013a,Bardeen2014,Amerongen2010,Schroter2015}.
Separation into the system and the bath parts is mostly formal as
the system-bath coupling has to be included to account for molecular
environment-induced decoherence and temperature effects, hence the
quantum dynamics penetrates into the bath, and bath also changes its
state.

When using wavefunction-based simulation approaches, it can be challenging
to maintain a precise representation of the bath as a constant temperature
thermostat, because energy exchange between the system and the bath
can alter thermal properties of the bath. Generally, a large number
of explicitly modelled QHO modes have to be included to minimize the
negative effects of thermal energy accumulation in the bath, but this
is numerically expensive. Therefore, one always has to balance between
the size of the model, accuracy of the chosen numerical method and
its numerical cost. Alternatively, one could numerically change the
wavefunction variables during its time evolution in a way as to prevent
accumulation of the thermal energy in the bath and to maintain it
at a desired temperature, i.e., perform thermalization.

It is challenging to accurately simulate dynamics of quantum systems
that exchange energy and (quasi-) particles with their surroundings,
i.e., of the open quantum systems \citep{Breuer2006,Weiss2012}, because
a numerical cost needed to propagate the corresponding dynamical equations
in time increases exponentially with the number of DOFs. The wavefunction
approach based on the multiple-Davydov $\text{D}_{2}$ trial wavefunction
($\mdd$ ansatz) \citep{Zhou2015b,Wang2016a,Zhou2016a,Chen2018,Zhao2021},
along with the time-dependent variational principle, has been shown
to be an excellent tool for accurately simulating the dynamics of
system-bath models \citep{Zhou2015b,Huang2017,Wang2017,Chen2019a,Jakucionis2020a,Wang2021a,MantasJakucionis2022,MantasJakucionis2022a}
and spectroscopic signals \citep{Sun2015a,Zhou2016,MantasJakucionis2022,MantasJakucionis2022a,Zhao2023}.
Despite relying on an adaptive, time-dependent state basis set, the
problem of rapidly growing numerical costs remain.

In a previous study, we proposed the thermalization algorithm \citep{Jakucionis2021}
to be used with the Davydov $\text{D}_{2}$ ansatz \citep{Sun2010c,Luo2010,Huang2017,Jakucionis2018d,Jakucionis2022,Davydov1979,Scott1991},
which restricts QHOs to their lowest uncertainty states -- coherent
states \citep{Zhang1990,Kais1990} with Gaussian wavepackets in their
coordinate-momentum phase space. We demonstrated how the thermalization
algorithm regulates the temperature of the environment and enables
the electronically excited molecular system to relax into its equilibrium
state at a given temperature \citep{Moix2012,Subas2012,Gelzinis2020}
even when using a reduced number of bath oscillators, which greatly
reduces numerical costs. The characteristics of the resulting equilibrium
state are essential for modeling fluorescence, excited state emission,
excited state absorption and other spectroscopic signals \citep{Mukamel1995}.
However, the $\dd$ ansatz is a crude approximation of the actual
system-bath model eigenstates and thus is unable to completely capture
electronic population relaxation dynamics \citep{Zhou2016}.

Meanwhile, the system-bath dynamics obtained using the multiple-Davydov
ansätze are consistent with the results from other state-of-the-art
methods, such as hierarchical equations of motion \citep{Zhou2016,Wang2016a,Chen2018},
quasiadiabatic path integral \citep{Wang2017}, and multi-configuration
time-dependent Hartree \citep{Chen2019a,Zeng2022}, even when the
number of bath oscillators is large. Due to the more complicated wavefunction
structure of the $\mdd$ ansatz, straightforward application of the
$\text{D}_{2}$ ansatz thermalization algorithm is not possible. In
this work, we extend the thermalization algorithm for the $\mdd$
ansatz by introducing an additional state projection algorithm and
adopting the coarse-grained scattering approximation.

In Section (\ref{sec:md2-therm-theory}) we describe the thermalization
algorithm for the $\text{mD}_{2}$ ansatz, and in Section (\ref{sec:Thermalized-fluorescence-spectra})
we provide a theoretical description of its application to simulating
the fluorescence spectra. Then, in Section (\ref{sec:Results}) we
demonstrate its capabilities by simulating excitation relaxation dynamics
of an H-type molecular aggregate and its fluorescence spectrum. Lastly,
in Section (\ref{sec:Discussion}) we discuss changes made to adapt
the thermalization algorithm of the $\text{D}_{2}$ ansatz for the
$\text{mD}_{2}$ ansatz.

\section{Thermalization of the $\text{mD}_{2}$ ansatz\label{sec:md2-therm-theory}}

We consider a molecular aggregate model, where each molecule $n=1,2,\ldots,N$
couples to its own \textit{local reservoir} $k=1,2,\ldots,N$, each
of which consists of $q=1,2,\ldots,Q$ QHO modes. The model is given
by the Hamiltonian $\hat{H}=\hat{H}_{\text{S}}+\hat{H}_{\text{B}}+\hat{H}_{\text{SB}}$
with the system, the bath and the system-bath coupling terms defined
as:
\begin{align}
\hat{H}_{\text{S}}= & \sum_{n}^{N}\varepsilon_{n}\hat{a}_{n}^{\dagger}\hat{a}_{n}+\sum_{n,m}^{n\neq m}J_{nm}\hat{a}_{n}^{\dagger}\hat{a}_{m},\label{eq:Hamil-S}\\
\hat{H}_{\text{B}}= & \sum_{k,q}^{N,Q}\omega_{kq}\hat{b}_{kq}^{\dagger}\hat{b}_{kq},\label{eq:Hamil-B}\\
\hat{H}_{\text{SB}}= & -\sum_{n}^{N}\hat{a}_{n}^{\dagger}\hat{a}_{n}\sum_{q}^{Q}\omega_{nq}g_{nq}\left(\hat{b}_{nq}^{\dagger}+\hat{b}_{nq}\right),\label{eq:Hamil-SB}
\end{align}
with the reduced Planck's constant set to $\hbar=1$. Here $\varepsilon_{n}$
is the $n$-th molecule electronic excitation energy, $J_{nm}$ denotes
the resonant coupling between the $n$-th and $m$-th molecules, $\omega_{nq}$
denotes the frequency of the $q$-th QHO in the $k$-th local reservoir,
while $g_{nq}$ is the coupling strength between the $q$-th oscillator
in the $n$-th local reservoir to the $n$-th molecule. The operators
$\hat{a}_{n}^{\dagger}$ and $\hat{a}_{n}$ represent the creation
and annihilation operators for electronic excitations, respectively,
while $\hat{b}_{nq}^{\dagger}$ and $\hat{b}_{nq}$ represent the
creation and annihilation bosonic operators for QHOs.

In addition, we implicitly couple the system-bath model to the secondary
bath characterized by a fixed temperature, $T.$ The coupling between
the secondary and primary baths occurs via the scattering events that
allow the system-bath model to exchange energy with the secondary
bath and thermalize local reservoirs as is described below.

A state of the system-bath model is given by the $\text{mD}_{2}$
wavefunction
\begin{equation}
|\Psi\left(t\right)\rangle=\sum_{i,n}^{M,N}\alpha_{i,n}\left(t\right)|n\rangle\otimes|\boldsymbol{\lambda}_{i}\left(t\right)\rangle,\label{eq:mD2}
\end{equation}
where $\alpha_{i,n}\left(t\right)$ is the $i$-th multiple complex
amplitude associated with a singly excited electronic state $|n\rangle$
localized on the $n$-th molecule, $|n\rangle=\hat{a}_{n}^{\dagger}|0\rangle_{el}$.
$|0\rangle_{el}$ is the electronic ground state. The complexity and
accuracy of the $\mdd$ ansatz can be adjusted by varying the multiplicity
number, $M$. The states of QHO modes are represented by multi-dimensional
coherent states
\begin{align}
|\boldsymbol{\lambda}_{i}\left(t\right)\rangle & =\exp\sum_{k,q}^{N,Q}\left(\lambda_{i,kq}\left(t\right)\hat{b}_{kq}^{\dagger}-\lambda_{i,kq}^{\star}\left(t\right)\hat{b}_{kq}\right)|0\rangle_{\text{vib}},
\end{align}
with $\lambda_{i,kq}\left(t\right)$ being the $i$-th multiple complex
displacement parameter and $|0\rangle_{\text{vib}}=\otimes_{k,q}|0\rangle_{k,q}$
is the global vibrational ground state of all QHOs.

The $\text{mD}_{2}$ wavefunction describes a state of the system-bath
model as a superposition of $M$ multi-dimensional coherent state
terms, which allows it to represent a wide range of system-bath model
states beyond the Born-Oppenheimer and Gaussian approximations.

The thermalization algorithm for the $\text{mD}_{2}$ ansatz is realized
by stochastic scattering events \citep{Plenio1998,Luoma2020} during
time evolution of the wavefunction. These events change momenta, $p_{kq}$,
of \textit{all} $q$-th QHO modes of the $k$-th local reservoir at
once. We assume that the scattering probability, $P_{k}\left(\theta,\tau_{\text{sc}}\right)$,
of $\theta$ scattering events occurring per time interval, $\tau_{\text{sc}}$,
with a scattering rate, $\nu_{k}$, is given by a Poisson distribution
\begin{equation}
P_{k}\left(\theta,\tau_{\text{sc}}\right)=\frac{1}{\theta!}\left(\tau_{\text{sc}}\nu_{k}\right)^{\theta}\text{e}^{-\tau_{\text{sc}}\nu_{k}}.\label{eq:Poiss}
\end{equation}
Numerically, Poisson statistics are realised by simulating Bernoulli
processes \citep{Kampen2007a,Bertsekas2008} in the limit of $\tau_{\text{sc}}\rightarrow0$,
while maintaining the condition that $\nu_{k}\tau_{\text{sc}}\ll1$.
To simulate the scattering events we divide wavefunction propagation
into equal time length, $\tau_{\text{sc}}$, intervals
\begin{equation}
t_{i}=\left(i\tau_{\text{sc}},\left(i+1\right)\tau_{\text{sc}}\right],\ i=0,1,\ldots.
\end{equation}
At the end of each interval, for each local reservoir, we flip a biased
coin with the probability $\nu_{k}\tau_{\text{sc}}$ of landing \textit{heads}
for all local reservours. If a $k$-th coin lands\textit{ heads,}
we change momenta of \emph{all oscillator modes of that $k$-th reservoir,}
otherwise, no changes are made. A list of scattering moments at which
the numerical simulation is paused to perform the scatterings can
be precomputed prior to starting the simulation by drawing probabilities
for all local reservoir and all time intervals $t_{i}$ from Eq. (\ref{eq:Poiss}).

We assume that during the scattering event the local bath, which experiences
the scattering, acquires thermal-equilibrium kinetic energy. Such
state is given by a single coherent state for one specific QHO. In
order to set the new momenta values of the scattered reservoir oscillator
modes, we first project the $\text{mD}_{2}$ wavefunction of Eq. (\ref{eq:mD2})
to its single-multiple Davydov $\text{D}_{2}$ form
\begin{equation}
|\psi\left(t\right)\rangle=\sum_{n}^{N}\beta_{n}\left(t\right)|n\rangle\otimes|\tilde{\boldsymbol{\lambda}}\left(t\right)\rangle,\label{eq:D2}
\end{equation}
where $\beta_{n}$ are the projected complex electronic amplitudes
and $|\tilde{\boldsymbol{\lambda}}\left(t\right)\rangle$ is the projected
multi-dimensional coherent state, which is defined later. This follows
the decoherence idea \citep{Schlosshauer2007}, where the macroscopic
environment perform a collapse of the wavefunction into a set of preferred
states, in our case, the electronic-vibrational states, $|n\rangle\otimes|\tilde{\boldsymbol{\lambda}}\left(t\right)\rangle$.
The projected complex electronic amplitudes are equal to
\begin{equation}
\beta_{n}\left(t\right)=\sum_{i}^{M}\alpha_{i,n}\left(t\right)\langle\tilde{\boldsymbol{\lambda}}\left(t\right)|\boldsymbol{\lambda}_{i}\left(t\right)\rangle,
\end{equation}
while the projected multi-dimensional coherent state
\begin{align}
|\tilde{\boldsymbol{\lambda}}\left(t\right)\rangle & =\exp\sum_{k,q}^{N,Q}\left(\tilde{\lambda}_{kq}\left(t\right)\hat{b}_{kq}^{\dagger}-\tilde{\lambda}_{kq}^{\star}\left(t\right)\hat{b}_{kq}\right)|0\rangle_{\text{vib}}
\end{align}
is defined in terms of the complex displacements
\begin{align}
\tilde{\lambda}_{kq}\left(t\right) & =\frac{1}{\sqrt{2}}\left(x_{kq}\left(t\right)+\text{i}p_{kq}\left(t\right)\right),\label{eq:scat-mode}
\end{align}
where $x_{kq}\left(t\right)$ and $p_{kq}\left(t\right)$ are QHO
coordinate and momentum expectation values 
\begin{align}
x_{kq}= & \frac{1}{\sqrt{2}}\sum_{i,j,n}^{M,M,N}\alpha_{i,n}^{\star}\alpha_{j,n}\langle\boldsymbol{\lambda}_{i}|\boldsymbol{\lambda}_{j}\rangle\sum_{k,q}^{N,Q}\left(\lambda_{i,kq}^{\star}+\lambda_{j,kq}\right),\\
p_{kq}= & \frac{\text{i}}{\sqrt{2}}\sum_{i,j,n}^{M,M,N}\alpha_{i,n}^{\star}\alpha_{j,n}\langle\boldsymbol{\lambda}_{i}|\boldsymbol{\lambda}_{j}\rangle\sum_{k,q}^{N,Q}\left(\lambda_{i,kq}^{\star}-\lambda_{j,kq}\right),
\end{align}
calculated from the $\text{mD}_{2}$ ansatz, where $\langle\boldsymbol{\lambda}_{i}|\boldsymbol{\lambda}_{j}\rangle$
is the overlap of two coherent states
\begin{equation}
\langle\boldsymbol{\lambda}_{i}|\boldsymbol{\lambda}_{j}\rangle=\exp\sum_{k,q}^{N,Q}\left(\lambda_{i,kq}^{\star}\lambda_{j,kq}-\frac{1}{2}\left(\left|\lambda_{i,kq}\right|^{2}+\left|\lambda_{j,kq}\right|^{2}\right)\right).
\end{equation}
This completes the projection operation of the $\text{mD}_{2}$ state,
given by Eq. (\ref{eq:mD2}), into its simplified $\text{D}_{2}$
form in Eq. (\ref{eq:D2}).

Once the projected wavefunction is deduced, we modify momenta of the
scattered oscillators by sampling the QHO diagonal density operator
distribution in the coherent state representation at temperature $T$,
known as the Glauber-Sudarshan distribution \citep{Glauber1963,Chorosajev2016c,Wang2017b,Xie2017}
\begin{equation}
\mathcal{P}\left(\tilde{\lambda}_{kq}\right)=\mathcal{Z}_{kq}^{-1}\exp\left(-\left|\tilde{\lambda}_{kq}\right|^{2}\left(\text{e}^{\frac{\omega_{kq}}{k_{\text{B}}T}}-1\right)\right).\label{eq:G-S}
\end{equation}
For scattered modes, we set momenta values in Eq. (\ref{eq:scat-mode})
to
\begin{equation}
p_{kq}\left(t\right)=\sqrt{2}\text{Im}\left(\tilde{\lambda}_{kq}^{\mathcal{P}}\right),
\end{equation}
where $\tilde{\lambda}_{kq}^{\mathcal{P}}$ is a sample drawn from
the Glauber-Sudarshan distribution. $\mathcal{Z}_{kq}^{-1}$ and $\omega_{kq}$
are partition functions and frequencies of QHO, respectively, while
$k_{\text{B}}$ is the Boltzmann constant. During the scattering events,
coordinates, $x_{kq}$, of both the scattered and non-scattered modes
remain unchanged. Notice, that the local baths, which do not experience
scattering, remain unaffected by the scattering of other modes.

Now that the wavefunction of the system-bath model after scattering
is known (given by Eq. (\ref{eq:D2})) we rewrite it in the $\text{mD}_{2}$
wavefunction form of Eq. (\ref{eq:mD2}) by populating amplitudes
and displacements of the first multiple, $i=1$, as
\begin{align}
\alpha_{1,n}\left(t\right) & =\beta_{n}\left(t\right),\\
\lambda_{1,kq}\left(t\right) & =\tilde{\lambda}_{kq}\left(t\right).
\end{align}
Amplitudes of the unpopulated multiples are set to $\alpha_{j=2,\ldots,M,n}\left(t\right)=0$,
while the unpopulated displacements are positioned in a layered hexagonal
pattern around the populated coherent state \citep{MantasJakucionis2022}
\begin{align}
\lambda_{j=2,\ldots,M,kq}\left(t\right) & =\lambda_{1,kq}\left(\tau\right)\nonumber \\
 & +\frac{1}{4}\left(1+\left\lfloor \beta\left(j\right)\right\rfloor \right)e^{\text{i}2\pi\left(\beta\left(j\right)+\frac{1}{12}\left\lfloor \beta\left(j\right)\right\rfloor \right)},
\end{align}
where $\beta\left(j\right)=\left(j-2\right)/6$ is the coordination
function and $\left\lfloor x\right\rfloor $ is the floor function.
The exact arrangement of displacements of the unpopulated multiples
is not critical, as long as the distance in the phase space to the
populated multiple coherent state is not too large, otherwise, the
initially unpopulated multiples will not contribute to further dynamics
\citep{Jakucionis2020a,Werther2020}.

Once the scattered $\mdd$ wavefunction is determined and the scattering
event is finalised, further simulation of $\text{mD}_{2}$ dynamics
according to equations of motion proceeds. This procedure generates
a stochastic wavefunction trajectory, where the system-bath model
at each time moment is described by a pure state, which is a single
member of a statistical ensemble \citep{Chorosajev2016c,Wang2017b}.
The thermalized model dynamics are obtained by averaging observables
over an ensemble of wavefunction trajectories, $\gamma$, which differ
by their: initial amplitudes, $\alpha_{i,n}\left(0\right)$, initial
coherent state displacements, $\lambda_{i,kq}\left(0\right)$, and
a sequence of scattering events. Ensemble averaging is performed in
a parallelized Monte Carlo scheme.

\section{Thermalized fluorescence spectra\label{sec:Thermalized-fluorescence-spectra}}

Wavefunction trajectories allows calculation of an arbitrary observable.
Calculation of equilibrium fluorescence spectrum requires to know
thermally equilibrated state of the excited model. The presented thermalization
procedure allows to obtain such state and calculate fluorescence spectrum.

In general, the frequency-domain spectrum of a quantum system can
be written as a Fourier transform
\begin{equation}
F\left(\omega\right)=\text{Re}\int_{0}^{\infty}\text{d}t\text{e}^{i\omega t}S\left(t\right),
\end{equation}
of the corresponding time-domain response function, $S\left(t\right)$.

The fluorescence (FL) response function, $S_{\text{fl}}\left(t\right)$,
is a specific case of the more general time-resolved fluorescence
(TRF) response function \citep{Mukamel1995,Balevicius2015c}
\begin{align}
S_{\text{trf}}\left(\tau_{\text{eq}},t\right) & =\frac{1}{\Gamma}\sum_{\gamma=1}^{\Gamma}\langle\Psi_{\text{G}}\left(0\right)|_{\gamma}\hat{\mu}_{-}\hat{\mathcal{V}}_{\text{E}}^{\dagger}\left(\tau_{\text{eq}}+t\right)\hat{\mu}_{+}\nonumber \\
 & \times\hat{\mathcal{V}}_{\text{G}}\left(t\right)\hat{\mu}_{-}\hat{\mathcal{V}}_{\text{E}}\left(\tau_{\text{eq}}\right)\hat{\mu}_{+}|\Psi_{\text{G}}\left(0\right)\rangle_{\gamma},\label{eq:trf-simple}
\end{align}
where $\hat{\mathcal{V}}_{\text{E}}$ and $\hat{\mathcal{V}}_{\text{G}}$
are the excited and ground state system-bath propagators
\begin{align}
\hat{\mathcal{V}}_{A}\left(t_{1}\right)|\Psi_{A}\left(t_{2}\right)\rangle & =|\Psi_{A}\left(t_{1}+t_{2}\right)\rangle,
\end{align}
while $\hat{\mu}_{+}=\sum_{n}^{N}\left(\boldsymbol{e}\cdot\boldsymbol{\mu}_{n}\right)\hat{a}_{n}^{\dagger}$
and $\hat{\mu}_{-}=\sum_{n}^{N}\left(\boldsymbol{e}\cdot\boldsymbol{\mu}_{n}\right)\hat{a}_{n}$
are the excitation creation and annihilation operators of the system
\citep{MantasJakucionis2022}, $\boldsymbol{\mu}_{n}$ is the electronic
transition dipole moment vector, $\boldsymbol{e}$ is the external
field polarization vector. $|\Psi_{\text{G}}\left(0\right)\rangle_{\gamma}$
is a model ground state with an initial condition of the $\gamma$-th
trajectory. The EOMs for propagating the $\mdd$ wavefunction, as
well as the approach to solving them, are described in detail in Refs.
\citep{Werther2020,MantasJakucionis2022}.

$S_{\text{trf}}\left(\tau_{\text{eq}},t\right)$ is a function of
two times: the equilibration time, $\tau_{\text{eq}}$, and the coherence
time, $t$. During the equilibration time, the system evolves in its
excited state and, due to the system-bath interaction, relaxes to
an equilibrium state. After this, during the coherence time, spontaneous
emission occurs.

We will apply thermalization during the equilibration time to facilitate
the relaxation of the system-bath model into the lowest energy equilibrium
state by removing excess thermal energy from local reservoirs. We
denote $\hat{\mathcal{G}}_{\text{E},\gamma}$ as the excited state
propagator $\hat{\mathcal{V}}_{\text{E}}$, but with the thermalization.
Then the thermalized TRF (tTRF) response function can be written as
\begin{align}
\tilde{S}_{\text{trf}}\left(\tau_{\text{eq}},t\right) & =\frac{1}{\Gamma}\sum_{\gamma=1}^{\Gamma}\langle\Psi_{\text{G}}\left(0\right)|_{\gamma}\hat{\mu}_{-}\hat{\mathcal{G}}_{\text{E},\gamma}^{\dagger}\left(\tau_{\text{eq}}\right)\hat{\mathcal{V}}_{\text{G}}^{\dagger}\left(t\right)\hat{\mu}_{+}\nonumber \\
 & \times\hat{\mathcal{V}}_{\text{G}}\left(t\right)\hat{\mu}_{-}\hat{\mathcal{G}}_{\text{E},\gamma}\left(\tau_{\text{eq}}\right)\hat{\mu}_{+}|\Psi_{\text{G}}\left(0\right)\rangle_{\gamma}.\label{eq:S_trf_g}
\end{align}

By considering the equilibration time to be long enough to reach the
equilibrium state of the system-bath model, we define the FL response
function to be
\begin{equation}
S_{\text{fl}}\left(t\right)=\lim_{\tau_{\text{eq}}\rightarrow\infty}S_{\text{trf}}\left(\tau_{\text{eq}},t\right),\label{eq:fl}
\end{equation}
and the thermalized fluorescence (tFL) response function as
\begin{equation}
\tilde{S}_{\text{fl}}\left(t\right)=\lim_{\tau_{\text{eq}}\rightarrow\infty}\tilde{S}_{\text{trf}}\left(\tau_{\text{eq}},t\right).\label{eq:tfl}
\end{equation}

The spectra obtained using the fluorescence response function without
and with thermalization will be compared in the next section. For
the numerical simulation, the required equilibration time interval
has to be deduced by increasing $\tau_{\text{eq}}$ until the resulting
fluorescence spectra converges.

\section{Results\label{sec:Results}}

To investigate the thermalization algorithm for the $\text{mD}_{2}$
ansatz, we will analyse the linear trimer model, which we previously
used to study thermalization of the $\text{D}_{2}$ ansatz \citep{Jakucionis2021}.
The model consists of $N=3$ coupled molecules, with excited state
energies $\varepsilon_{n}$ being equal to $0,\ 250,\ 500\ \text{cm}^{-1}$,
forming an energy funnel. The nearest neighbour couplings are set
to $J_{1,2}=J_{2,3}=100\ \text{cm}^{-1}$, $J_{3,1}=0$. The electronic
dipole moment vectors of molecules are $\boldsymbol{\mu}_{n}=(1,0,0)$
in Cartesian coordinate system. This classifies the trimer as the
H-type molecular aggregate \citep{Hestand2018}.

QHOs of local molecular reservoirs are characterized by the super-Ohmic
\citep{Kell2013} spectral density function $C''\left(\omega\right)=\omega\left(\omega/\omega_{c}\right)^{s-1}\exp\left(-\omega/\omega_{c}\right)$
with an order parameter $s=2$ and a cut-off frequency $\omega_{\text{c}}=100\ \cm$.
The QHO frequencies are $\omega_{kq}=\omega_{0}+\left(q-1\right)\Delta\omega$,
where the frequency off-set is $\omega_{0}=0.01\ \text{cm}^{-1}$.
The reorganization energy of each local reservoir is $\Lambda_{k}=\sum_{q}\omega_{kq}g_{kq}^{2}=100\ \cm$.
The scattering time step-size is set to $\tau_{\text{sc}}=0.01\ \text{ps}$.
Finally, the ensemble consists of $900$ wavefunctions trajectories,
which we found to be sufficient to obtain the converged model dynamics.
The $\text{mD}_{2}$ ansatz multiplicity $M=5$ is used as the results
with a higher multiplicity quantitatively match the $M=5$ case.

We will be considering three bath models: the \textit{dense bath}
model, where the spectral density function $C''\left(\omega\right)$
is discretized into $Q=75$ oscillators per local reservoir with a
step-size of $\Delta\omega=10\ \text{cm}^{-1}$; the \textit{sparse
bath} model, where the number of modes is reduced by a factor of 5
to just $Q=15$ oscillators per local reservoir with $\Delta\omega=50\ \text{cm}^{-1}$;
and the \textit{sparse bath with thermalization} model, where $C''\left(\omega\right)$
is discretized according to the sparse bath model and thermalization
is used.

In the absence of the bath, the system has three single-excitation
stationary exciton states with energies: $E_{1}^{\text{exc}}\approx-37.23\ \cm$,
$E_{2}^{\text{exc}}=250\ \cm$, $E_{3}^{\text{exc}}\approx537.23\ \cm$,
satisfying the time-independent Schrödinger equation
\begin{equation}
\hat{H}_{\text{S}}\Phi_{n}^{\text{exc}}=E_{n}^{\text{exc}}\Phi_{n}^{\text{exc}},
\end{equation}
with the system Hamiltonian given by Eq. (\ref{eq:Hamil-S}). The
exciton eigenstates \citep{Valkunas2013a,Amerongen2010}, $\Phi_{n}^{\text{exc}}$,
have their excitations delocalized over multiple molecules \citep{Chorosajev2016c}.
Therefore, it is convenient to analyse molecular aggregate excitation
relaxation dynamics in terms of excitons as they are natural quasi-particles
of the aggregate. We denote the probability of finding the aggregate
in its $n$-th excitonic state as the population, given by
\begin{equation}
\rho_{n}^{\text{exc}}\left(t\right)=\text{\ensuremath{\sum_{k,l,i,j}}}\left(\Phi_{k}^{\text{exc}}\right)_{n}^{\star}\left\langle \alpha_{i,k}^{\star}\left(t\right)\alpha_{j,l}\left(t\right)S_{i,j}\left(t\right)\right\rangle _{\text{th}}\left(\Phi_{l}^{\text{exc}}\right)_{n},
\end{equation}
where $\left\langle \ldots\right\rangle _{\text{th}}$ is the averaging
over an ensemble of wavefunction trajectories.

Using $\text{mD}_{2}$ ansatz we have

\begin{figure}
\includegraphics[width=1\columnwidth]{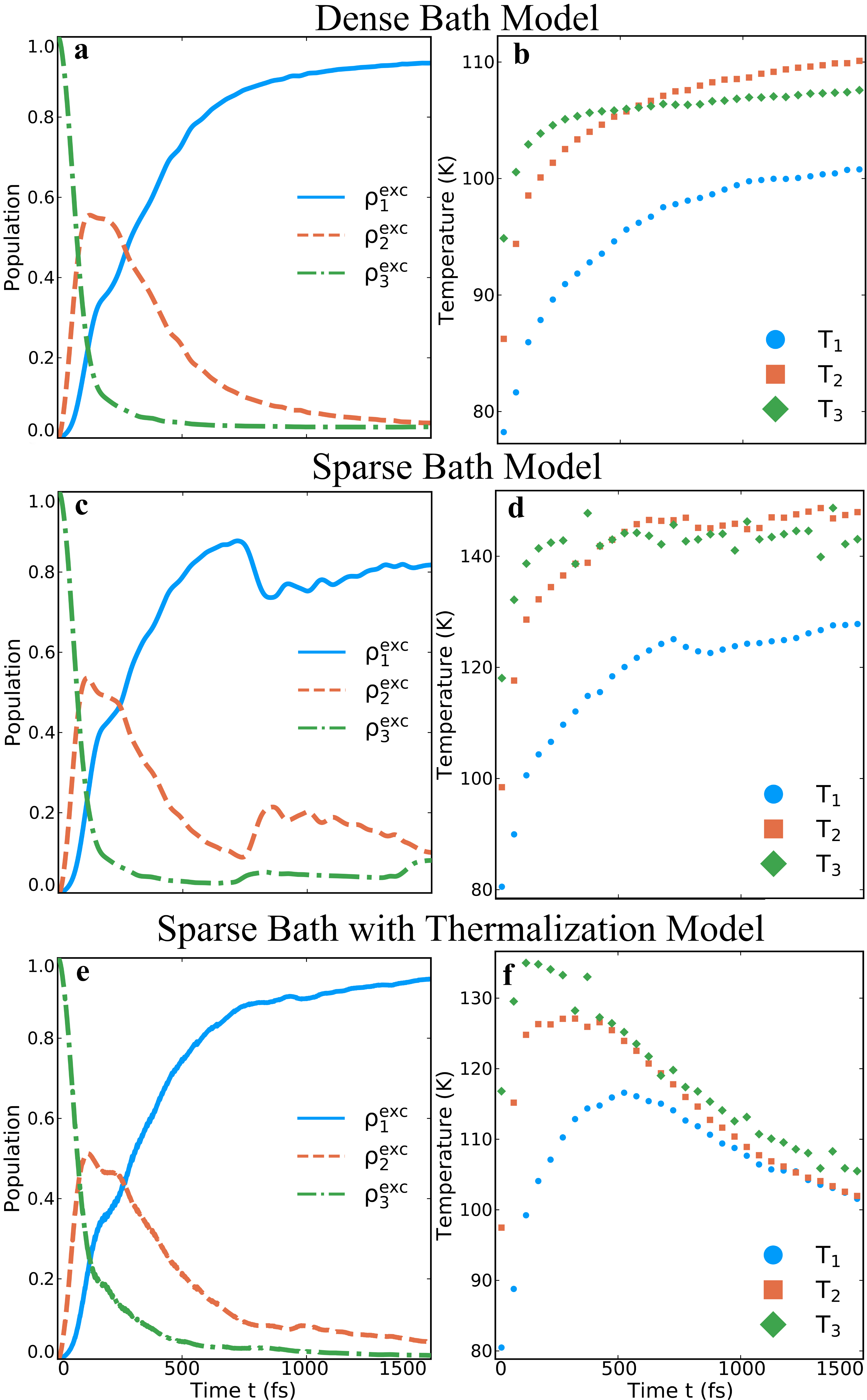}

\caption{The exciton state populations, $\rho_{n}^{\text{exc}}\left(t\right)$,
and the average temperatures, $T_{k}\left(t\right)$, of local reservoirs
of the trimer with (a, b) the dense bath model, (c, d) the sparse
bath model and (e, f) the sparse bath model with thermalization.\label{fig:pt}}
\end{figure}

First, we study the electronic excitation dynamics. The initial excitonic
state populations correspond to the optically excited highest energy
state: $\rho_{3}^{\text{exc}}=1$, $\rho_{1,2}^{\text{exc}}=0$, while
the initial QHO displacements, $\lambda_{i,kq}\left(0\right)$, are
sampled from the Glauber-Sudarshan distribution in Eq. (\ref{eq:G-S})
to account for the initial temperatures of $T_{k}=77\ \text{K}$.

In Fig. (\ref{fig:pt}) we display the trimer model exciton state
populations $\rho_{n}^{\text{exc}}\left(t\right)$ and average temperatures
\citep{Abramavicius2018c} $T_{k}\left(t\right)$ of local reservoirs
for all three bath models. The aggregate environment causes dephasing
between excitonic states and induces irreversible population relaxation
\citep{Mukamel1995,Valkunas2013a}. The population dynamics of the
dense bath models exhibits a sequential relaxation from the initially
populated highest energy excitonic state to the lowest energy state
via the intermediate state. Eventually, the population distribution
reaches the equilibrium state. The majority of the excitation energy
is transferred to oscillators of local reservoirs. We observe an increase
of temperatures \citep{Scarani2002,Abramavicius2018c,Chin2013a} due
to the finite number of oscillators in local reservoirs. An infinite
number of oscillators would have to be included to maintain temperature
constant at the initial value. The initial rapid rise in temperature
is due to oscillator reorganization in the aggregate's electronic
excited state manifold, while the following slow rise is due to energy
transfer from the system to local reservoirs.

In the sparse bath model, we observe that if the number of vibrational
modes is reduced, the population dynamics become skewed due to insufficiently
dense representation of the spectral density function. Furthermore,
the temperature increase is higher than compared to the dense bath
model, which further changes the characteristics of the resulting
equilibrium state.

When the thermalization algorithm is applied to the sparse bath model
with a scattering rate $\nu_{k}=1.25\ \text{ps}^{-1}$, the population
dynamics are restored and qualitatively match those of the dense model.
Although the initial temperatures of the local reservoirs exceed those
of the dense bath model, they gradually decrease due to thermalization,
and this rate can be adjusted by changing the scattering rate.
\begin{figure}
\includegraphics[width=1\columnwidth]{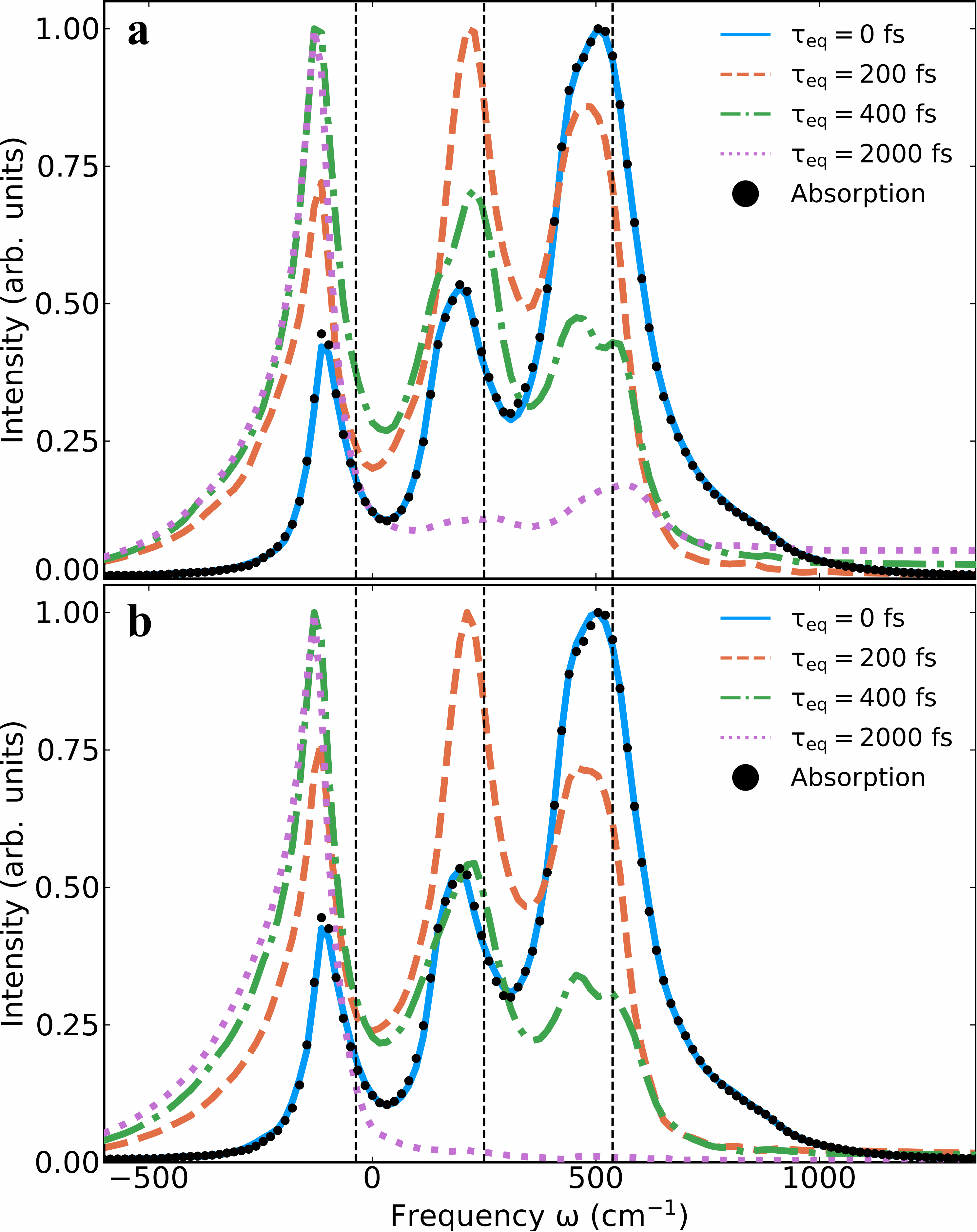}

\caption{(a) the TRF and (b) tTRF spectra of the trimer with the dense bath
model, simulated with an increasing equilibration time $\tau_{\text{eq}}$.
The absorption spectrum is also shown. Vertical dashed lines show
energies $E^{\text{exc}}$ of the excitonic states. \label{fig:fl_time}}
\end{figure}

Next, we turn our attention to simulating the FL spectrum of the linear
trimer with the dense bath model with scattering rate $\nu_{k}=1\ \text{ps}^{-1}$.
The initial excitonic state population distribution is now calculated
in terms of the system-field interaction, as described in Ref. \citep{MantasJakucionis2022}.

In Fig. (\ref{fig:fl_time}) we compare the TRF and tTRF spectra with
increasing equilibration times $\tau_{\text{eq}}$. When $\tau_{\text{eq}}=0$,
both the TRF and tTRF spectra are equivalent and exactly match the
absorption spectrum, which consists of three peaks due to transition
involving the combined excitonic-vibrational (vibronic) states and
can not be regarded as purely excitonic. For reference, vertical dotted
lines indicate energies $E^{\text{exc}}$ of excitonic states. These
do not match the three peak energies exactly due to the system being
coupled to the environment.

By allowing equilibration to occur, $\tau_{\text{eq}}>0$, both the
TRF and tTRF spectra show peak intensity shift towards lower energies
as excitation relaxes towards the equilibrated state during the equilibration
time. After equilibrating for $\tau_{\text{eq}}=2\ \text{ps}$, we
find that both spectra have converged and do not change with longer
$\tau_{\text{eq}}$. Therefore, the TRF and tTRF spectra at $\tau_{\text{eq}}=2\ \text{ps}$
can be considered as the FL and tFL spectra of the trimer model as
defined in Eqs. (\ref{eq:fl}), (\ref{eq:tfl}).

Both spectra exhibit their highest intensities at the energies of
the lowest vibronic states. However, the FL spectrum also has considerable
intensities at energies of the intermediate and highest vibronic states.
Surprisingly, the higher energy peak is more intense than the intermediate
peak. The tFL spectrum intensities at these energies are negligible,
which indicate that the thermalization allows the trimer model to
reach a lower energy equilibrium state, which is no longer hindered
by the excess thermal energy accumulation in QHOs of local reservoirs.

\begin{figure}
\includegraphics[width=1\columnwidth]{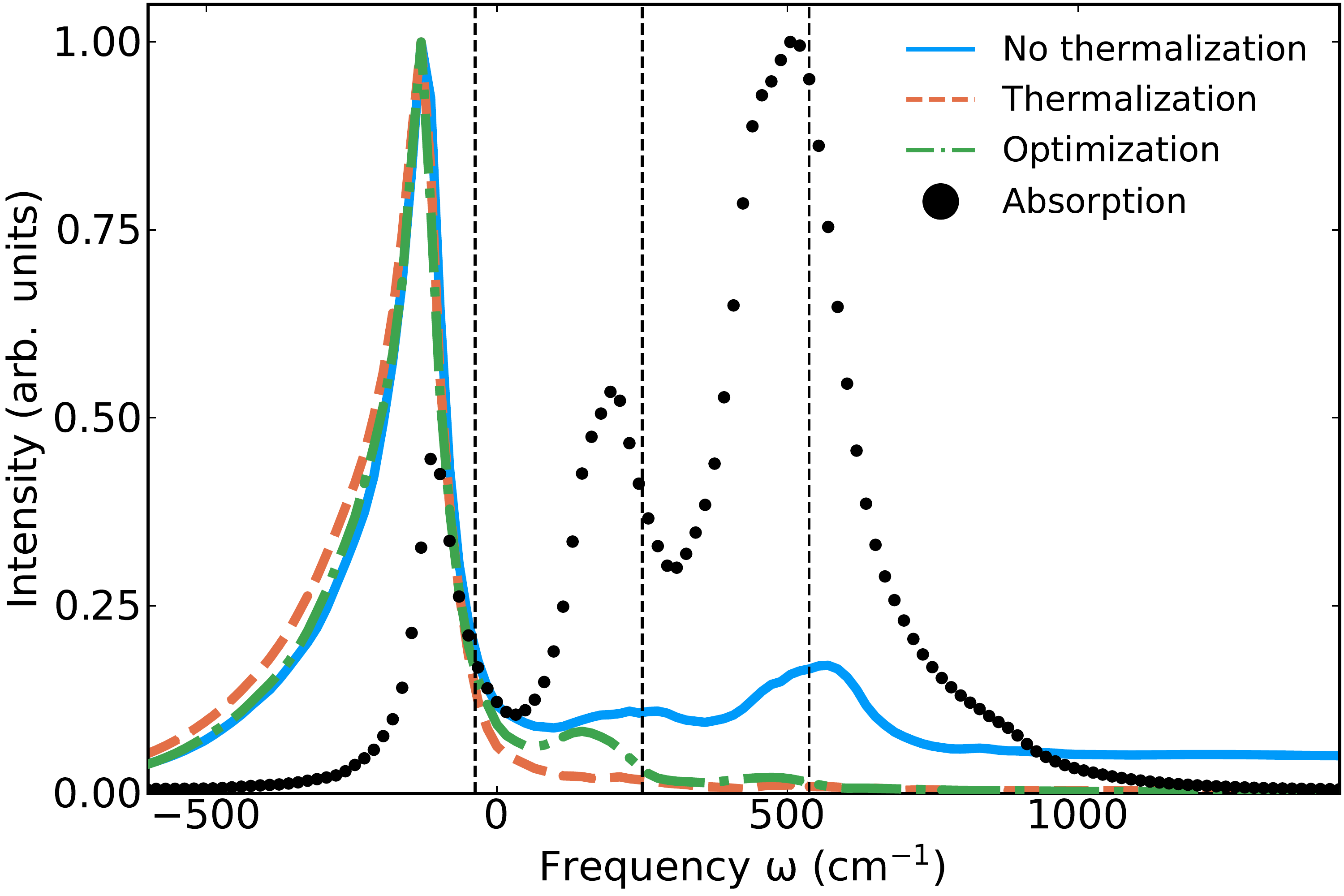}

\caption{The fluorescence spectra comparison of the trimer with the dense bath
model obtained without thermalization, with thermalization and using
the optimization approach. The equilibration time is $\tau_{\text{eq}}=2\ \text{ps}$.
The absorption spectrum is also shown. Vertical dashed lines show
energies $E^{\text{exc}}$ of the excitonic states.\label{fig:fl_comparison}}
\end{figure}

In Fig. (\ref{fig:fl_comparison}), we also compare the obtained FL
and tFL spectra with the FL spectrum simulated using a previously
proposed excited state numerical optimization approach \citep{MantasJakucionis2022a,mogensen2018optim,Zhan2009}.
It relies on finding the model's lowest energy excitonic state in
terms of the $\text{mD}_{2}$ ansatz parameters and then applying
thermal fluctuations to effectively generate the model in a lowest
energy equilibrium state at the temperature of $77\ \text{K}$. We
see that all three methods produce a similar lowest vibronic peak,
but the tFL spectrum has a higher intensity tail towards the low energy
side and almost no intensities at energies of the intermediate and
the highest vibronic states, while the FL spectrum simulated using
the optimization approach has a small intensity at the energy of the
intermediate vibronic states. The optimization approach spectrum more
closely resembles that of the thermalized model than the non-thermalized
spectrum.

\section{Discussion\label{sec:Discussion}}

Starting from an arbitrary non-equilibrium initial condition a closed
quantum system will not equilibrate due to energy conservation. Thermalization
procedure is necessary to guarantee proper thermal equilibrium in
the long run for \emph{all }bath oscillators. This requires introducing
the concept of primary and secondary baths. In our model the primary
bath is a part of explicit quantum DOFs, while the secondary bath
is a thermal reservour with infinite thermal capacity, i. e. it keeps
constant temperature in any energy exchange process. In this case,
the secondary bath cannot be described by mechanical equations --
only statistical or thermodynamical conepts apply. Our statistical
algorithm performs energy exchange between the primary and secondary
baths using the statistical scattering idea: the primary bath state
is being reset to the thermally equilibrated state, thus giving up
excess energy to or drawing additional energy from the secondary bath.
This is a major extension of the explicit quantum TDVP theory --
the extended model covers a broader range of phenomena: local heating
and cooling, as well as bath oscillator dynamic localization, which
are not available in the standard TDVP theory.

In order to adapt the $\dd$ ansatz thermalization algorithm for the
$\mdd$ ansatz, several extensions were made. During the time evolution
of the system-bath model, the $\mdd$ ansatz multiples become correlated,
leading to a non-Gaussian bath wavefunction. It becomes impossible
to represent a new Gaussian wavefunction of scattered QHO modes, sampled
from Eq. (\ref{eq:G-S}), without changing the wavefunction of all
the rest non-scattered oscillators at the same time. Therefore, we
chose to project the $\mdd$ into the $\dd$ ansatz whenever scatterings
occurs, allowing to correctly represent the newly sampled Gaussian
wavefunction of scattered oscillators. This idea requires the consideration
of a few aspects.

The projected $\dd$ wavefunction accurately maintains \textit{average}
coordinates and momenta of the $\mdd$ ansatz QHO states, while variances
and higher-order moments become affected. This causes variation of
excitation relaxation dynamics compared to the standard $\mdd$ ansatz.
However, system-bath models mostly rely only on the \textit{linear}
coupling between the system and average coordinates of QHO modes,
therefore, as seen in Fig. (\ref{fig:pt}), the discrepancy is minimal.
The higher-order couplings become necessary when anharmonic vibrational
modes or changes to their frequencies upon excitation are considered
\citep{Chorosajev2017a,MantasJakucionis2022a}.

To maintain the close correspondence to the standard $\mdd$ ansatz
the projection should not occur too often. This is because it takes
time for the wavefunction after scattering to again become correlated
between its many multiples, i.e., to take advantage of the unpopulated
$\mdd$ ansatz multiples after projection. If the repopulation time
is shorter than the time between projection operations, the model
population dynamics become similar to those of the $\dd$ ansatz,
even though the $\mdd$ ansatz is being used. The average time interval
between projection operations is determined by the scattering rate
$\nu_{k}$, a property of the physical system, while the scattering
time $\tau_{\text{sc}}$ is a parameter of the model and must be as
small as necessary to ensure the Bernoulli-to-Poisson statistics transition
condition, $\nu_{k}\tau_{\text{sc}}\ll1$.

To increase the average time between projection operations, we adopt
a coarser scattering approach for the $\mdd$ ansatz as compared to
the $\dd$ ansatz. Instead of considering scattering events of individual
oscillators, we consider events, where all oscillators of certain
local reservoirs are scattered at once, requiring only a single projection
operation to scatter many oscillators at once. This approach allows
the $\mdd$ ansatz to continue utilizing all its multiples for the
improved accuracy over the $\dd$ ansatz, while reducing the number
of explicitly modelled oscillators needed to maintain local reservoirs'
temperatures close to initial values, thereby reducing the numerical
cost.

Using the $\mdd$ ansatz to simulate population dynamics of the trimer
model, it took an average of 166 minutes per trajectory using the
dense bath model, but only 1.3 minutes using the sparse bath model
and 2 minutes using the sparse bath model with thermalization. The
computational overhead of thermalization is small compared to the
overall time savings when switching from using the dense bath to the
sparse bath. The numerical cost reduction is also greater for the
$\mdd$ ansatz than for the $\dd$ ansatz in Ref. \citep{Jakucionis2021},
because the $\mdd$ ansatz EOMs constitute an \textit{implicit} system
of differential equations, which require a more involved, two-step
numerical approach to find a solution \citep{MantasJakucionis2022,Werther2020}.
By considering fewer oscillators in each local reservoir, simulations
of the dynamics and spectroscopic signals of aggregates made up of
more molecules becomes possible.

Computing a single trajectory of the tTRF response function in Eq.
(\ref{eq:S_trf_g}) with an equilibration time of $\tau_{\text{eq}}=2\ \text{ps}$
took an average of 79 minutes. The previously proposed optimization
method \citep{MantasJakucionis2022a} for simulating FL spectra does
not require propagation during the equilibration time interval of
the TRF response function and has to be computed only once, but it
took 193 minutes. In general, we find that the computation of tTRF
is more reliable and numerically stable. The optimization approach
struggles to consistently find the lowest energy excitonic state of
the model due to its heuristic nature, requiring many attempts to
find the solution and eventually having to choose the lowest energy
one. This is particularly apparent when a wide range of oscillator
frequencies are included.

For elementary system-bath models without Hamiltonian parameter disorder,
the optimization approach can be a good starting point for FL spectra
simulation. However, a more accurate spectra most likely will be obtained
using the tTRF approach. For models with Hamiltonian disorder, e.g.,
static molecule excitation energy disorder \citep{Abramavicius2003,Eisfeld2010b,Rancova2017},
the optimization approach would require finding model's lowest energy
excitonic state for each realization of the Hamiltonian, negating
its advantage of having to perform optimization procedure only once.

In conclusion, the presented thermalization algorithm for the numerically
exact $\mdd$ ansatz allows to reduce numerical cost of system-bath
model simulations by having to explicitly include fewer bath oscillators,
while maintaining correspondence with the exact relaxation dynamics.
The thermalization algorithm efficiently controls molecular heating
effects due to the reduced number of oscillators. Furthermore, the
application of thermalization to simulation of fluorescence spectra
demonstrates lower computation time, greater numerically stability
and higher accuracy compared to the numerical optimization approach.

\section*{Conflicts of interest}

There are no conflicts of interest to declare.
\begin{acknowledgments}
We thank the Research Council of Lithuania for financial support (grant
No: S-MIP-23-48). Computations were performed on resources at the
High Performance Computing Center, \textquoteleft \textquoteleft HPC
Sauletekis\textquoteright \textquoteright{} in Vilnius University
Faculty of Physics.\bibliographystyle{apsrev4-2}
\bibliography{mD2-therm}
\end{acknowledgments}

\end{document}